\documentstyle[sprocl]{article}
\def\be{\begin{equation}}
\def\ee{\end{equation}}
\def\bea{\begin{eqnarray}}
\def\eea{\end{eqnarray}}

\begin{document}
\title{SUPERSYMMETRIC U(1) AT THE TEV SCALE}
\author{E. KEITH}
\address{Center for Particle Physics, University of Texas\\
Austin, TX 78712, USA}
\author{ERNEST MA}
\address{Department of Physics, University of California\\
Riverside, CA 92521, USA}
\maketitle\abstracts{If there exists an arbitrary supersymmetric U(1) 
gauge factor at the TeV scale, under which the two Higgs superfields 
$H_{1,2}$ of the standard model are nontrivial, and if there is also 
a singlet superfield $S$ such that the $H_1 H_2 S$ term is allowed in 
the superpotential, then the structure of the two-doublet Higgs sector 
at the electroweak scale is more general than that of the MSSM (Minimal 
Supersymmetric Standard Model).  Under further assumptions of grand 
unification and universal soft supersymmetry breaking terms, the scale 
of U(1) breaking is related to the parameter $\tan \beta \equiv v_2/v_1$.}

\section{Two Doublets and a Singlet}

Assume an arbitrary supersymmetric U(1) gauge factor at the TeV scale.  
Call it $U(1)_X$.  Then under $SU(3)_C \times SU(2)_L \times U(1)_Y \times 
U(1)_X$, consider two doublet and one singlet Higgs superfields transforming 
as follows:~\cite{1}
\begin{eqnarray}
H_1 &\sim& (1, 2, -1/2; -a), \\ H_2 &\sim& (1, 2, 1/2; -1 + a), \\ 
S &\sim& (1, 1, 0; 1).
\end{eqnarray}
The superpotential of this model is then given by
\begin{equation}
W = f H_1 H_2 S + ...
\end{equation}
This has the advantage that the $\mu H_1 H_2$ term in the MSSM (Minimal 
Supersymmetric Standard Model) is replaced by $\mu = f \langle S \rangle$, 
which is a possible solution of the so-called $\mu$ problem.  (Since the 
$H_1 H_2$ term is allowed by supersymmetry in the MSSM, there is no natural 
understanding as to why $\mu$ should not be very much larger than the scale of 
supersymmetry breaking.  On the other hand, with $\langle S \rangle$ at the 
TeV scale and $f$ a typical Yukawa coupling, it is natural for $\mu$ to 
be at the electroweak scale, leading to the possibility that all physical 
members of the two Higgs doublets are at the 100 GeV scale.)

In the scalar sector, let $H_1$ be represented by $\tilde \Phi_1 = 
(\bar \phi_1^0, -\phi_1^-)$ and $H_2$ by $\Phi_2 = (\phi_2^+, \phi_2^0)$, 
and $S$ by $\chi^0$, then $\langle \chi^0 \rangle = u$ breaks only $U(1)_X$, 
and $\langle \phi_{1,2}^0 \rangle = v_{1,2}$ breaks $SU(2)_L \times U(1)_Y$ 
to $U(1)_Q$.  The contribution of the $f H_1 H_2 S$ term in the superpotential 
to the Higgs potential is given by
\begin{equation}
V_F = f^2 [ (\Phi_1^\dagger \Phi_2)(\Phi_2^\dagger \Phi_1) + (\Phi_1^\dagger 
\Phi_1 + \Phi_2^\dagger \Phi_2) \bar \chi \chi ],
\end{equation}
and the gauge contribution is
\begin{eqnarray}
V_D &=& {1 \over 8} g_2^2 [ (\Phi_1^\dagger \Phi_1)^2 + (\Phi_2^\dagger 
\Phi_2)^2 + 2 (\Phi_1^\dagger \Phi_1)(\Phi_2^\dagger \Phi_2) - 
4(\Phi_1^\dagger \Phi_2)(\Phi_2^\dagger \Phi_1) ] \nonumber \\ &+& 
{1 \over 8} g_1^2 [ \Phi_1^\dagger \Phi_1 - \Phi_2^\dagger \Phi_2 ]^2 \nonumber 
\\ &+& {1 \over 2} g_x^2 [ -a \Phi_1^\dagger \Phi_1 - (1-a) \Phi_2^\dagger 
\Phi_2 + \bar \chi \chi ]^2.
\end{eqnarray}
The supersymmetry breaking terms are contained in
\begin{equation}
V_{soft} = \mu_1^2 \Phi_1^\dagger \Phi_1 + \mu_2^2 \Phi_2^\dagger \Phi_2 + 
m^2 \bar \chi \chi + (f A_f \Phi_1^\dagger \Phi_2 \chi + h.c.).
\end{equation}

First, we assume $u \sim M_{SUSY} \sim$ TeV.  This is natural because 
$U(1)_X$ cannot be broken without also breaking the supersymmetry.  Of course, 
it is also possible~\cite{2} to have $u <$ TeV.  Second, we assume that at 
the 100 GeV scale, there are just the two Higgs doublets.  This requires 
$f A_f u = m_{12}^2 << ({\rm TeV})^2$.

\section{Reduced Higgs Potential}

With $\langle \chi \rangle = u$, the scalar field $\sqrt 2 Re \chi$ is 
physical and has a mass given by $m^2 = 2 g_x^2 u^2$.  The cubic 
$(\Phi_1^\dagger \Phi_1) \sqrt Re \chi$ coupling is then $\sqrt 2 u (f^2 
- g_x^2 a)$.  Hence the effective quartic $(\Phi_1^\dagger \Phi_1)^2$ 
coupling is given by
\begin{eqnarray}
\lambda_1 &=& {1 \over 4} (g_1^2 + g_2^2) + g_x^2 a^2 - {{2 u^2 (f^2 - 
g_x^2 a)^2} \over {2 g_x^2 u^2}} \nonumber \\ &=& {1 \over 4} (g_1^2 + 
g_2^2) + 2 a f^2 - {f^4 \over g_x^2}.
\end{eqnarray}
Similarly, the other effective quartic scalar couplings are as follows:
\begin{eqnarray}
\lambda_2 &=& {1 \over 4} (g_1^2 + g_2^2) + 2(1-a) f^2 - {f^4 \over g_x^2}, 
\\ \lambda_3 &=& -{1 \over 4} g_1^2 + {1 \over 4} g_2^2 + f^2 - {f^4 \over 
g_x^2}, \\ \lambda_4 &=& -{1 \over 2} g_2^2 + f^2.
\end{eqnarray}
Note that in the limit $f=0$, we recover the Higgs structure of the MSSM.

Let $\tan \beta \equiv v_2/v_1$ and $v \equiv (v_1^2 + v_2^2)^{1/2}$, then
\begin{equation}
(m_h^2)_{max} = 2 v^2 [\lambda_1 \cos^4 \beta + \lambda_2 \sin^4 \beta + 
2(\lambda_3 + \lambda_4) \sin^2 \beta \cos^2 \beta] + \epsilon,
\end{equation}
where the radiative correction due to the $t$ quark and squarks is given by
\begin{equation}
\epsilon \simeq {{3 g_2^2 m_t^4} \over {8 \pi^2 M_W^2}} \ln \left( 1 + 
{\tilde m^2 \over m_t^2} \right).
\end{equation}
Hence the existence of an extra supersymmetric U(1) gauge factor at the 
TeV scale implies
\begin{equation}
(m_h^2)_{max} = M_Z^2 \cos^2 2 \beta + \epsilon + {f^2 \over {\sqrt 2 G_F}} 
\left[ A - {f^2 \over g_x^2} \right],
\end{equation}
where
\begin{equation}
A = {3 \over 2} + (2a-1) \cos 2 \beta - {1 \over 2} \cos^2 2 \beta.
\end{equation}
If $A > 0$, the MSSM bound can be exceeded.  However, for a given 
$g_x^2$, $f^2$ is still bounded from the requirement that the reduced 
Higgs potential for the two doublets be bounded from below.  Hence 
$(m_h^2)_{max}$ is bounded.  See Figure 1 of Ref.~1.  For specific models 
derivable from $E_6$, the upper bound is raised to no higher than about 
146 GeV as compared to 128 GeV in the MSSM.  For $g_x^2 = 0.5$, the 
absolute upper bound is about 190 GeV.

\section{Models Based on $E_6$} 

Consider the sequential reduction of $E_6$:
\begin{eqnarray}
E_6 &\rightarrow& SO(10)~[\times U(1)_\psi], \\ SO(10) &\rightarrow& 
SU(5)~[\times U(1)_\chi], \\ SU(5) &\rightarrow& SU(3)_C \times SU(2)_L 
~[\times U(1)_Y].
\end{eqnarray}
Assuming that a single extra U(1) survives down to the TeV energy scale, 
it is generally given by a linear combination of $U(1)_\psi$ and 
$U(1)_\chi$ which we call $U(1)_\alpha$.  Under the maximal subgroup 
$SU(3)_C \times SU(3)_L \times SU(3)_R$, the fundamental representation 
of $E_6$ is given by
\begin{equation}
{\bf 27} = (3, 3, 1) + (3^*, 1, 3^*) + (1, 3^*, 3).
\end{equation}
Under the subgroup $SU(5) \times U(1)_\psi \times U(1)_\chi$, we then have
\begin{eqnarray}
{\bf 27} = (10; 1, -1) [(u,d),u^c,e^c] + (5^*; 1, 3) [d^c,(\nu_e,e)] + 
(1; 1, -5) [N] &~& \nonumber \\ + (5; -2, 2) [h,(E^c,N_E^c)] + 
(5^*; -2, -2) [h^c,(\nu_E,E)] + (1; 4, 0) [S], &~&
\end{eqnarray}
where the U(1) charges refer to $2 \sqrt 6 Q_\psi$ and $2 \sqrt {10} Q_\chi$. 
Note that the known quarks and leptons are contained in $(10; 1, -1)$ and 
$(5^*; 1, 3)$, and the two Higgs scalar doublets are represented by 
$(\nu_E,E)$ and $(E^c,N_E^c)$.  Let
\begin{equation}
Q_\alpha = Q_\psi \cos \alpha - Q_\chi \sin \alpha,
\end{equation}
then the $\eta$-model~\cite{3} is obtained with $\tan \alpha = \sqrt {3/5}$ 
and we have
\begin{equation}
{\bf 27} = (10;2) + (5^*;-1) + (1;5) + (5;-4) + (5^*;-1) + (1;5),
\end{equation}
where $2 \sqrt {15} Q_\eta$ is denoted.  Hence $a = 1/5$ and $g_x^2 = 
(25/36) g_1^2$ for this model.  If we take $\tan \alpha = -1/\sqrt {15}$, 
we get the $N$-model~\cite{4} with
\begin{equation}
{\bf 27} = (10;1) + (5^*;2) + (1;0) + (5;-2) + (5^*;-3) + (1;5),
\end{equation}
where $2 \sqrt {10} Q_N$ is denoted.  Here $a = 3/5$ and $g_x^2 = (25/24) 
g_1^2$.  As a last example, in the exotic left-right model~\cite{5}, 
$a = \tan^2 \theta_W$ and
\begin{equation}
g_x^2 = {{(g_1^2 + g_2^2) (1-\sin^2 \theta_W)^2} \over {4(1-2 \sin^2 
\theta_W)}}.
\end{equation}

\section{Supersymmetric Scalar Masses}

As a reasonable and predictive procedure, we will adopt the common hypothesis 
that soft supersymmetry-breaking operators appear at the grand-unification 
scale as the result of a hidden sector which is linked to the observable 
sector only through gravity.  Hence these terms will be assumed to be 
universal, {\it i.e.} of the same magnitude for all fields.  Consider now 
the masses of the supersymmetric scalar partners of the quarks and leptons:
\begin{equation}
m_B^2 = m_0^2 + m_R^2 + m_F^2 + m_D^2,
\end{equation}
where $m_0$ is a universal soft supersymmetry-breaking mass, $m_R^2$ is a 
correction generated by the renormalization-group equations running from the 
grand-unification scale down to the TeV scale, $m_F$ is the explicit mass of 
the fermion partner, and $m_D^2$ is a term induced by gauge symmetry 
breaking with rank reduction and can be expressed in terms of the gauge-boson 
masses.  In the MSSM, $m_D^2$ is of order $M_Z^2$ and does not change $m_B$ 
significantly.  In the $U(1)_\alpha$-extended model, $m_D^2$ is of order 
$M_{Z'}^2$, and will affect $m_B$ in a nontrivial way.  The contributions 
to $m_D^2$ from $U(1)_\alpha$ are
\begin{eqnarray}
\Delta m_D^2 (10;1,-1) &=& {1 \over 8} M_{Z'}^2 \left( 1 + \sqrt {3 \over 5} 
\tan \alpha \right), \\ \Delta m_D^2 (5^*;1,3) &=& {1 \over 8} M_{Z'}^2 
\left( 1 - 3 \sqrt {3 \over 5} \tan \alpha \right), \\ \Delta m_D^2 (1; 1;-5) 
&=& {1 \over 8} M_{Z'}^2 \left( 1 + \sqrt {15} \tan \alpha \right), \\ 
\Delta M_D^2 (5;-2,2) &=& -{1 \over 4} M_{Z'}^2 \left( 1 + \sqrt {3 \over 5} 
\tan \alpha \right), \\ \Delta m_D^2 (5^*;-2,-2) &=& -{1 \over 4} M_{Z'}^2 
\left( 1 - \sqrt {3 \over 5} \tan \alpha \right), \\ \Delta m_D^2 (1;4,0) 
&=& {1 \over 2} M_{Z'}^2.
\end{eqnarray}
Thus it is actually possible~\cite{6} for exotic scalar quarks and leptons 
to be lighter than the ordinary ones.  Furthermore, since these masses are 
also present in the Higgs potential, the contributions of $\Delta m_D^2$ 
due to $U(1)_\alpha$ are essential in constraining its parameters, {\it i.e.} 
$m_A$ and $\tan \beta$.

\section{Matching of Parameters at the TeV Scale}

In the two-doublet Higgs potential, the soft terms are given by
\begin{equation}
V_{soft} = m_1^2 \Phi_1^\dagger \Phi_1 + m_2^2 \Phi_2^\dagger \Phi_2 + 
m_{12}^2 (\Phi_1^\dagger \Phi_2 + \Phi_2^\dagger \Phi_1).
\end{equation}
In the $U(1)_\alpha$-extended model, they are related to $M_Z$ and the 
pseudoscalar mass $m_A$ as follows:
\begin{eqnarray}
m_{12}^2 &=& - m_A^2 \sin \beta \cos \beta, \\ m_1^2 &=& m_A^2 \sin^2 \beta 
-{1 \over 2} M_Z^2 \cos 2 \beta \nonumber \\ &-& {{2 f^2} \over g_Z^2} 
M_Z^2 \left[ 2 \sin^2 \beta  + \left( 1 - \sqrt {3 \over 5} \tan \alpha 
\right) \cos^2 \beta - {{3 f^2} \over {2 \cos^2 \alpha g_\alpha^2}} \right], 
\\ m_2^2 &=& m_A^2 \cos^2 \beta + {1 \over 2} M_Z^2 \cos 2 \beta \nonumber \\ 
&-& {{2 f^2} \over g_Z^2} M_Z^2 \left[ 2 \cos^2 \beta + \left( 1 + \sqrt 
{3 \over 5} \tan \alpha \right) \sin^2 \beta - {{3 f^2} \over {2 \cos^2 
\alpha g_\alpha^2}} \right].
\end{eqnarray}
In the limit $f=0$ in the above, we recover the well-known results of the 
MSSM.

On the other hand, Eqs.~(7) and (25) tell us that
\begin{eqnarray}
m_{12}^2 &=& f A_f u, \\ m_1^2 &=& m_0^2 + m_{R1}^2 + f^2 u^2 - {1 \over 4} 
\left( 1 - \sqrt {3 \over 5} \tan \alpha \right) M_{Z'}^2, \\ m_2^2 &=& 
m_0^2 + m_{R2}^2 + f^2 u^2 - {1 \over 4} \left( 1 + \sqrt {3 \over 5} 
\tan \alpha \right) M_{Z'}^2,
\end{eqnarray}
where $M_{Z'}^2 = (4/3) \cos^2 \alpha g_\alpha^2 u^2$ and $m_{R2}^2$ 
differs from $m_{R1}^2$ in that the former contains the contribution from 
the $t$ Yukawa coupling and the latter does not.  Given a particular 
$U(1)_\alpha$ from $E_6$, we can start at the grand-unification scale 
with $m_0$, $A_0$, and $m_{1/2}$, then for a given value of $f$, the 
matching of Eqs.~(36) to (38) with Eqs.~(33) to (35) will allow us to 
derive $u$ and $\tan \beta$.

In our approach~\cite{1}, we assume that the term $f' h h^c S$ in the 
superpotential is important enough to drive $m_\chi^2$ in Eq.~(7) negative, 
so that $U(1)_\alpha$ is broken with
\begin{equation}
M_{Z'} = \sqrt {-2 m_\chi^2} = {2 \over \sqrt 3} g_\alpha \cos \alpha |u|,
\end{equation}
where $g_\alpha$ is assumed equal approximately to $\sqrt {5/3} g_1$. 
The mass of the exotic quark $h$ is then given by $f' |u|$.  As mentioned 
already at the end of Sec.~1, we assume also that $f A_f$ to be small 
compared to $u$.  In a different approach~\cite{2}, $f'$ is assumed zero, 
but $f A_f$ is taken to be rather large.  In that case, $v_{1,2}$ and $u$ 
cannot be separated into two different scales.  Also, there is no solution 
with universal soft supersymmetry-breaking terms.

In Figures 4 to 9 of Ref.~1, we show typical solutions of $\tan \beta$ 
and $|u|$ for various inputs of $m_0$, $A_0$, the gluino mass, and $f$. 
We find $M_{Z'}$ to be around 1 TeV, $m_h$ around 2 TeV, and $\tan \beta$ 
around 4.  These solutions are much more constrained than the ones in the 
MSSM because the free parameter $m_{12}^2 = B \mu$ is now replaced by 
$f A_f u$.

\section{Conclusion}

An extra supersymmetric $U(1)_\alpha$ gauge factor from $E_6$ is a good 
possibility at the TeV energy scale.  Its existence implies that the 
two-doublet Higgs structure at around 100 GeV will be observably different 
from that of the MSSM.  Supersymmetric scalar masses are also 
very different because of the large $\Delta m_D^2$ contributions. 
The $U(1)_\alpha$ breaking scale and the well-known parameter $\tan \beta 
\equiv v_2/v_1$ are closely related.

\section*{Acknowledgments} The presenter of this talk (E.M.) thanks Goran 
Senjanovic and Alexei Smirnov for the organization of this stimulating 
conference.  This work was supported in part by the U.~S.~Department of 
Energy under Grant No.~DE-FG03-94ER40837.

\end{document}